\documentclass[apj]{emulateapj}

\usepackage{natbib}
\usepackage{color}
\usepackage{graphicx}
\bibpunct{(}{)}{;}{a}{}{,}

\begin{document}

   \title{
Metallicity inhomogeneities in local star-forming galaxies\\
as sign of recent metal-poor gas accretion
}
   \author{
          	J.~S\'anchez~Almeida\altaffilmark{1,2},
          	A.~B.~Morales-Luis\altaffilmark{1,2},
        	C.~Mu\~noz-Tu\~n\'on\altaffilmark{1,2},\\ 
                D.~M.~Elmegreen\altaffilmark{3}, 
                B.~G.~Elmegreen\altaffilmark{4},
                and
         	J.~M\'endez-Abreu\altaffilmark{1,2,5}
          }
\altaffiltext{1}{Instituto de Astrof\'\i sica de Canarias, E-38205 La Laguna,
Tenerife, Spain}
\altaffiltext{2}{Departamento de Astrof\'\i sica, Universidad de La Laguna,
Tenerife, Spain}
\altaffiltext{3}{Department of Physics and Astronomy, Vassar College, 
Poughkeepsie, NY 12604, USA}
\altaffiltext{4}{IBM Research Division, T.J. Watson Research Center, Yorktown Heights, 
NY 10598, USA}
\altaffiltext{5}{
School of Physics and Astronomy, University of St Andrews, North Haugh, St Andrews, KY16 9SS, UK}
\email{jos@iac.es, abml@iac.es, cmt@iac.es, elmegreen@vassar.edu, bge@us.ibm.com,
jma20@st-andrews.ac.uk}
\begin{abstract}
We measure the oxygen metallicity 
of the ionized gas
along the major axis 
of seven dwarf star-forming  galaxies.  
Two of them,  {\tt SDSSJ1647+21} and {\tt SDSSJ2238+14}, 
show $\simeq0.5$\,dex 
metallicity decrements  in inner regions with 
enhanced star-formation  activity. 
This behavior is similar to the metallicity drop  observed in a 
number of local tadpole galaxies by \citeauthor{2013ApJ...767...74S}, 
and interpreted as showing early stages of assembling in disk galaxies, 
with the star formation sustained by external metal-poor gas
accretion.
The agreement with tadpoles has several implications: (1) it proves that
galaxies other than the local tadpoles present the same unusual  
metallicity pattern. (2) Our metallicity inhomogeneities 
were inferred using the direct method, thus discarding 
systematic  errors usually attributed to other methods. 
(3) Taken together with the tadpole data, our findings suggest 
a threshold around one tenth the solar value for the metallicity 
drops to show up.
Although galaxies with clear metallicity drops are rare, the
physical mechanism responsible for them
may sustain a  significant part of the star-formation activity in the 
local Universe. We argue that the star-formation dependence of
the mass-metallicity relationship, as well as other general 
properties followed by most local disk galaxies, 
are naturally  interpreted as side effects of pristine 
gas infall.
Alternatives to the metal poor gas accretion are 
examined too.
\end{abstract}



\section{Introduction}\label{introduction}

There are two major modes of galaxy formation, as inferred from 
cosmological numerical simulations 
\citep[e.g.,][]{2012RAA....12..917S,2006MNRAS.368....2D}.
At large redshifts major mergers play the dominant role,
where galaxies of similar masses merge to form larger agregates.
As the universe evolves, a second mechanism takes over.
The proto-galaxies grow by  accretion of external flows of 
pristine gas, that penetrate the dark matter halo 
and hit and heat a pre-existing elementary disk.
Cosmological simulations predict this cold-flow buildup to be the 
main mode of galaxy formation 
\citep{2009Natur.457..451D,2012ApJ...745...11G}, 
and the incoming gas is expected to form giant clumps that
spiral in and merge into a central spheroid 
\citep{1999ApJ...514...77N,2008ApJ...687...59G,2008ApJ...688...67E},
or just create thick disks that evolve by secular processes 
\citep[][]{2009Natur.457..451D,2012MNRAS.426..690B}. 

Observational evidence for this cold-flow accretion mode 
comes from the decrease of metallicity associated with
internal star-formation regions in high redshift disk galaxies
 \citep{2010Natur.467..811C}. Such localized metallicity drops 
in the inner disk cannot be explained in any other obvious way 
but the accretion of  external metal poor gas --  
secular evolution produces 
disks with a metallicity decreasing inside-out 
\citep[e.g.,][]{1988MNRAS.235..633V,2011ARA&A..49..301V,2012ApJ...745...66M},  
in sharp contrast with these observations.
The same kind of metallicity deficit associated 
with  bright star-forming regions has also been observed in a
particular type of local galaxies with tadpole morphology
\citep[][]{2013ApJ...767...74S}. Their images show a 
large star-forming 
clump at one end and a long diffuse region to one side. 
This asymmetric  morphology is fairly common at high 
redshift  
\citep{2007ApJ...658..763E,
2010ApJ...722.1895E, 
2006ApJ...639..724S,
2006NewAR..50..821W}
but rare in  the local Universe
\citep{2012ApJ...750...95E},
where it turns out to be associated with extremely metal 
poor galaxies, and so, with chemically primitive objects
\citep[][]{2008A&A...491..113P,2011ApJ...743...77M,2013A&A...558A..18F}.
These facts  were used by \citet{2012ApJ...750...95E} and 
\citet{2013ApJ...767...74S} to 
conjecture that local tadpole galaxies are disks in early stages of
assembling.
Metallicity drops associated  star-forming regions 
have also been observed in few other local targets, 
including a gamma ray burst host galaxy  \citep{2011ApJ...739...23L} 
and a Blue Compact Dwarf~(BCD) galaxy \citep{2010ApJ...715..656W}. 
They are interpreted in terms of redistribution of centrally 
generated metals,  with strong galactic winds and 
subsequent fallback, but not as cold-flow accretion events.

Here we analyze the spatial variation of metallicity in a set  of 
BCD galaxies with intense starbursts 
and having a range of metallicities from 2/3 to 1/20 the solar 
value.
The  purpose of the work is twofold. First, to see whether the 
metallicity inhomogeneities observed in tadpoles are also present 
in other local targets different from the original sample 
(Sect.~\ref{observations}). Second, and equally relevant,
to check if the metallicity variations remain when 
the metallicities are estimated via the direct 
method.
Thus we can discard a systematic error
in the strong-line method employed by  
\citet{2013ApJ...767...74S} to infer
the abundance inhomogeneities
\citep[e.g.,][]{2010IAUS..262...93S}.

The result of our analysis confirms that, at least in two  
targets, there are metallicity drops associated with intense 
starbursts. These drops are not present in the objects
of larger metallicity. We use this fact to conjecture that a 
minimum metallicity around 1/10 the solar value is needed 
for the metallicity decrements to be observed. 
Galaxies with metallicities below this one tenth 
threshold are usually referred to as 
extremely metal-poor \citep[XMP; e.g.,][]{2000A&ARv..10....1K}.

One might interpret the rarity of galaxies with 
metallicity drops as evidence against systematic cold-flow 
accretion in the local Universe. Thus the few observed decrements 
would represent  vestiges of a physical process
common early on, but now almost inoperative. 
However, several independent observations 
suggest that star-formation triggered by accretion  
of metal-poor (perhaps pristine) gas may be a process more common 
than anticipated.
Several of those evidences are put forward and discussed in
detail in Section~\ref{ubiquity}, all of them involving global
properties of large numbers of galaxies.
The most conspicuous one refers to the so-called 
mass-metallicity relation. It has been recently found 
\citep{
2010MNRAS.408.2115M,
2010A&A...521L..53L,
2013A&A...549A..25P,
2013ApJ...765..140A}
that  for galaxies of the same mass, their current 
star-formation rate (SFR) is anti-correlated  
with their ionized gas metallicity. No contrived explanation is required 
if the two parameters are physically connected, as if 
the infall of metal-poor gas feeds and triggers the star 
formation in these galaxies
\citep{2012ApJ...750..142B,2012MNRAS.421...98D}.

The paper is organized as follows: Section~\ref{observations}
describes observations and reduction. Metallicity estimates
are outlined in Section~\ref{metallicities}.
The resulting gradients and inhomogeneities 
are analyzed in Section~\ref{analysis}.
Potential observational biases and alternatives to the 
metal poor gas accretion are examined in Sect.~\ref{rule_out}.
Observational evidences for grand-scale gas inflows triggering
star-formation in the 
local Universe are presented and discussed in Section~\ref{ubiquity}.
The implications of our work are  considered in Sect.~\ref{discussion}.

%
\section{Observation and data analysis}\label{observations}

The seven galaxies used in this study are listed in Table~\ref{the_table}.
Even though their spectra were originally obtained with a different 
purpose\footnote{Specifically, for checking the metallicity of XMP
candidates selected from SDSS/DR6 as BCDs having 
negligible  [NII]$\lambda$6583\,\AA\ \citep[Sect.~2.2 in ][]{2008ApJ...685..194S}. 
The absence of this line is a signature of low metallicity 
\citep[e.g.,][]{2002MNRAS.330...69D,2011ApJ...743...77M},
but most of the candidates 
from \citet{2008ApJ...685..194S}
lack [NII]$\lambda$6583\,\AA\
due to an artifact of the reduction pipeline,
that removed [NII] together with an overlapping telluric line.
Thus they present a range of metallicities.},
they turned out to be ideal for our work. Their long-slit 
spectra provide spatial resolution within the targets,
with a spectral coverage enough to detect
all the lines required for oxygen abundance analysis using 
the direct method. The targets cover a wide range 
metallicities, from 2/3 to 1/20 the solar metallicity
\citep[see Table~\ref{the_table}, 
with $12+\log({\rm O/H})_\odot=8.69$ 
as measured by][]{2009ARA&A..47..481A}. 
In addition, the galaxies form stars actively,  in the sense that the 
current starburst is much larger than the average SFR during the galaxy 
lifetime
\citep[
assumed to be similar to the age of Universe $t_0$, since
the galaxies presumedly contain old stellar populations -- 
see, e.g., 
][]
{1996A&AS..120..207P,2006ApJ...651..861C,2012ApJ...756..163S}. 
The time-scale $t_\star$  to produce their stellar masses
${\rm M}_\star$ at the current star formation  rate (SFR), 
\begin{equation}
t_\star={\rm M}_\star/{\rm SFR},
\end{equation}
is typically much smaller than one Gyr  (see Table~\ref{the_table}), and so
much smaller than $t_0$ ($\simeq$14\,Gyr).
The SFRs and stellar masses  in Table~\ref{the_table} use  SDSS
H$\alpha$ fluxes, colors and magnitudes together with the prescriptions in  
\citet{1998ARA&A..36..189K} and  \citet{2012ApJ...750...95E}, and 
the mass-to-light ratios in \citet{2001ApJ...550..212B}.

All long-slit spectra were taken with the spectrograph ISIS of 
the  4.2\,m William Hershel Telescope (WHT) operated in the 
Roque de los Muchachos 
Observatory\footnote{{\tt http://www.ing.iac.es/astronomy/telescopes/wht/}}. 
The dual beam, red and blue, covers in a single
exposure  from $\lambda$3600\,\AA\
to 8000\,\AA . The ISIS@WHT setup includes intermediate 
gratings which, 
after a 2$\times$2 binning of the CCD, 
provide 1.7\,\AA\,pix$^{-1}$ (blue) and  
1.9\,\AA\,pix$^{-1}$ (red) equivalent to  
0\farcs40\,pix$^{-1}$ (blue) and 0\farcs44\,pix$^{-1}$ (red).
We use a slit 1\arcsec\ wide, which limits the angular
resolution, and also sets the spectral resolution to some
4.2\,\AA\ both in the red and the blue arms. This resolution
suffices to measure the fluxes of the relevant emission lines
[O{\sc iii}]$\lambda\lambda$4363,4959,5007\,\AA ,
[O{\sc ii}]$\lambda\lambda$3727,7319,7330\,\AA ,
H$\beta$, H$\alpha$, [N{\sc ii}]$\lambda$6584\,\AA ,
and [S{\sc ii}]$\lambda\lambda$6717,6731\,\AA .
The observations were carried out in two campaigns 
(Jan 31, 2009, and July 15, 2010), both with fair-to-good  
seeing from 1\farcs3 and 0\farcs5. We integrated 4000\,s
on target.
Some of the objects show an elongated 
morphology (Table~\ref{the_table}), and then  the slit was oriented 
along the major axis. Otherwise, the slit followed the parallactic 
angle.

\begin{figure}
\includegraphics[width=0.5\textwidth]{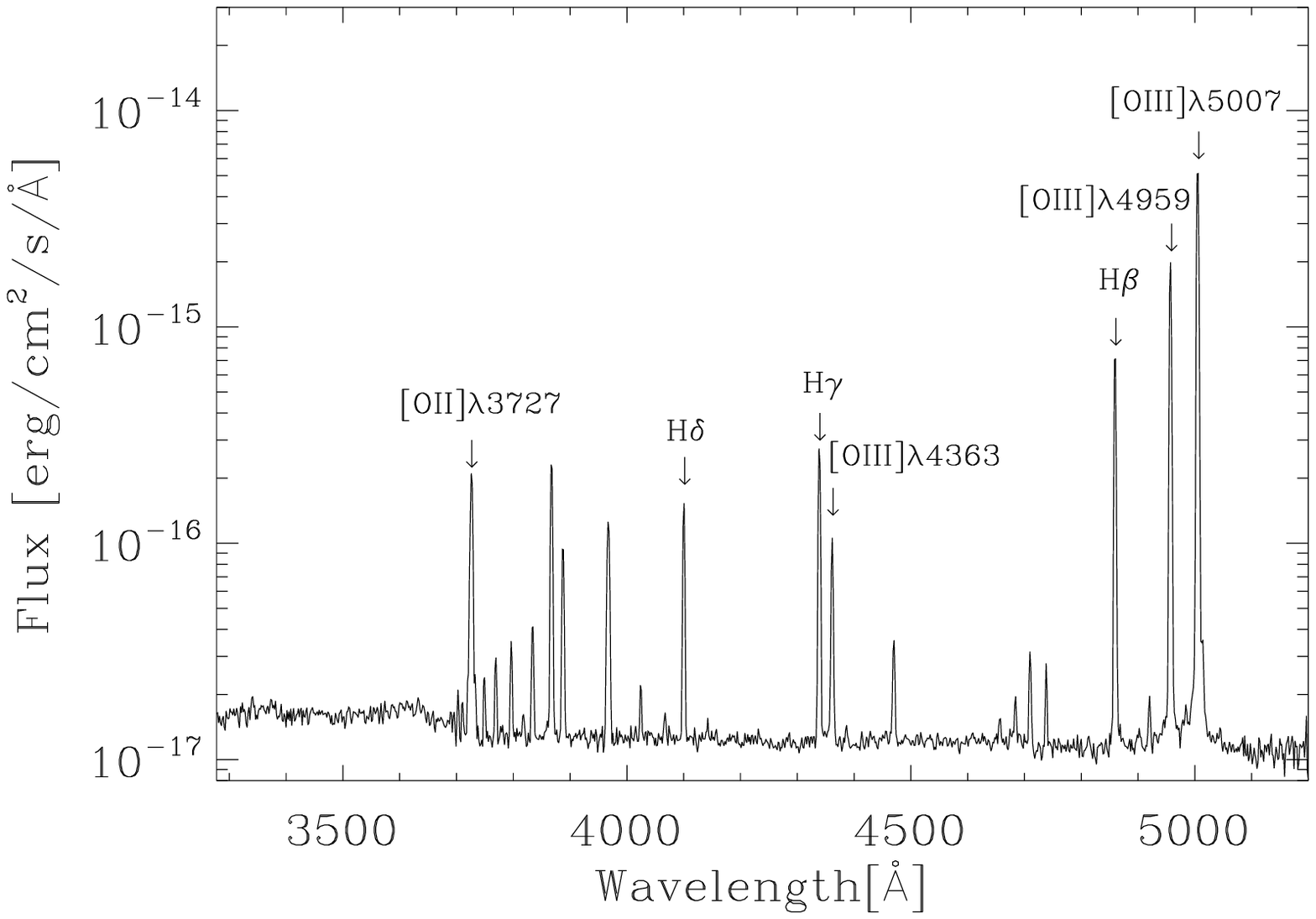}
\includegraphics[width=0.5\textwidth]{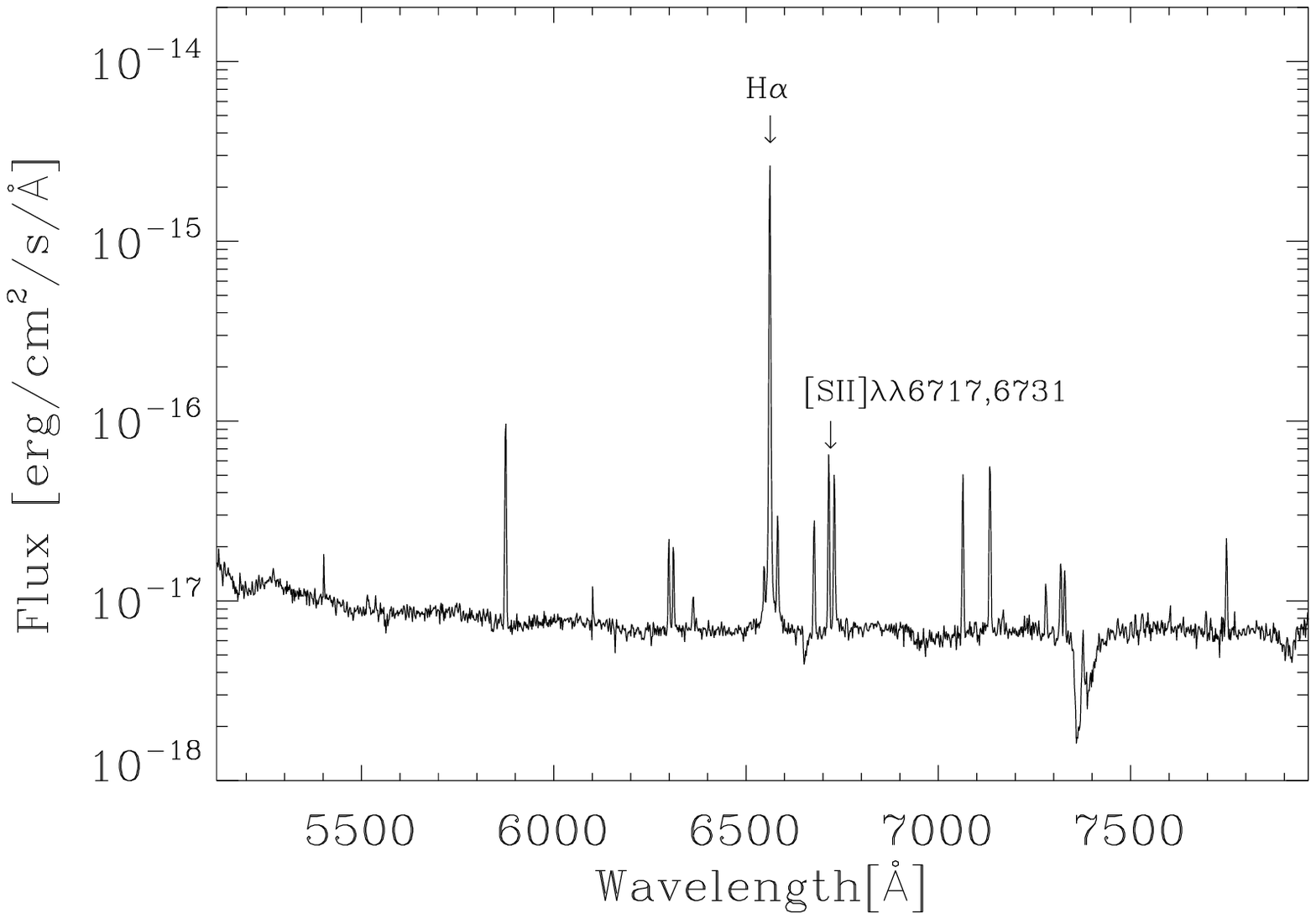}
\caption{
Example of one of the spectra used in the work, with
the main emission lines included.  
It corresponds to the brightest knot of  {\tt J1509+37}. 
The spectra of the two arms of the spectrograph are shown 
in different panels -- the upper and the lower panels 
correspond the blue and the red arms, respectively.
Fluxes are given in a logarithmic scale to show faint lines. 
}
\label{my_example}
\end{figure}

The reduction procedure included  standard bias and flatfield 
corrections, cosmic ray elimination, absolute flux calibration, 
as well as removal of sky emission lines.
Spectral and spatial directions were not exactly 
perpendicular on the CCD, and we also correct
for this effect. The
spectra were aligned so that each 
column corresponds to a single position on the sky.
Thus the different columns are extracted and analyzed 
independently in the paper,
with each spectrum representing 0\farcs44 on the galaxy.
After these manipulations, 
the signal-to-noise ratio (S/N)  in H$\alpha$ turns out to be between
1000 and 300 from the center to the outskirts of a typical 
galaxy. The critical line needed for electron 
temperature determination, [OIII]$\lambda$4363\,\AA , 
is much fainter than H$\alpha$, but it still reaches a S/N 
up to 70 in the brighter regions. As we explain below, 
${\rm S/N}\rightarrow 0$
(this line disappears) when the metallicity becomes large.  
Figure~\ref{my_example} contains an example of one of these
fully reduced spectra, specifically, the brightest knot of {\tt J1509+37}
(Table~\ref{the_table}). 

\subsection{Metallicity Determination}\label{metallicities}

We determine the oxygen abundance using the direct method 
\citep[e.g.,][]{1995PASP..107..896S,2004cmpe.conf..115S}, 
following the prescription by \citet[][]{2008MNRAS.383..209H}, which
includes employing the Balmer decrement to correct for internal reddening.
Electron densities were calculated using the ratio of fluxes
[S{\sc ii}]$\lambda$6717\AA/[S{\sc ii}]$\lambda$6731\AA.
The electron temperature of [O{\sc iii}]
was derived from the ratio 
([O{\sc iii}]$\lambda$4959\AA+[O{\sc iii}]$\lambda$5007\AA)/[O{\sc iii}]$\lambda$4363\AA. 
The ratio   [O{\sc ii}]$\lambda$3727\AA/([O{\sc ii}]$\lambda$7319\AA +[O{\sc ii}]$\lambda$7330\AA ) 
was used to measure the 
electron temperature of [O{\sc ii}] or, when this line was not available,
 calculated using the relationship between [O{\sc ii}]
and [O{\sc iii}] temperatures worked out by \citet{2003MNRAS.346..105P} .
We use throughout
the manuscript the term {\em modified direct method}
to describe this approach to [O{\sc ii}] temperature estimate.
%
%
Finally, the oxygen metallicity is computed by adding up the 
contribution of all oxygen ionization states up to O$^{2+}$.  
The errors in the oxygen abundances were
computed in a Monte-Carlo simulation, by randomly 
modifying the fluxes of the emission lines according to the noise 
of the observed spectra as measured in their continua
and scaled up to account for the photon
noise \citep[e.g.,][Sect.~3.1]{2003MNRAS.346..105P}.
The abundances are computed from 500 realizations of the noise, 
and the standard deviation of the resulting O/H 
are quoted as error bar. In a second error estimate, we
repeated the Monte-Carlo exercise assuming the noise in continuum
to be three times the observed one. 

In addition to the direct method, in order to compare it
with the metallicities and metallicity variations found by
 \citet{2013ApJ...767...74S}, 
we also estimate the oxygen abundance using the 
ratio [NII]$\lambda$6583\,\AA\ to H$\alpha$. It is the so-called
N2-method as proposed by
\citet{2002MNRAS.330...69D}, and we use it 
in the calibration by \citet{2009MNRAS.398..949P}.

%
%
%
%
%

%
%
%
\begin{deluxetable*}{cccccccccc}
\tabletypesize{\scriptsize}
\tablewidth{0pt}
\tablecaption{Global parameters of the galaxies}
\tablehead{
\colhead{Name\tablenotemark{a}}&
\colhead{$12+\log({\rm O/H})$\tablenotemark{b}}&
\colhead{$g$\tablenotemark{c}}&
\colhead{$\log {\rm M}_\star$\tablenotemark{d}}&
\colhead{SFR\tablenotemark{e}}&
\colhead{${\rm M}_\star/{\rm SFR}$\tablenotemark{f}}&
\colhead{Redshift}&
\colhead{Morphology}&
\colhead{$\Delta\log({\rm O/H})$?}\\
&&&
$[{\rm M}_\odot]$&
$[{\rm M}_\odot {\rm yr}^{-1}]$&
$[$Gyr$]$&
$\times 10^2$&
}
\startdata
SDSSJ083713.13+360350.4& $8.54\pm 0.01$ &17.4&  9.14&0.68 &  2.0&3.31&single knot&no\\   
SDSSJ094254.27+340411.8& $7.79\pm 0.05$ &19.1&  7.34&0.33&  0.068&2.25&single knot&unclear\\
SDSSJ100348.65+450457.7& $7.89\pm 0.01$ &17.6&  7.60&0.15&  0.27&0.92&single knot&no\\   
SDSSJ150934.17+373146.1& $7.80\pm  0.01$ &17.3&  7.51&3.62&  0.009& 3.25&cometary&no\\   
SDSSJ164710.66+210514.5& $8.11\pm 0.03$ &16.9&  7.73&0.43&  0.12&0.91&cometary&yes\\   
SDSSJ223831.12+140029.7& $7.43\pm 0.01$ &18.7&  7.55&0.51&  0.070&2.06&two-knots&yes\\   
SDSSJ230210.00+004938.8& $7.75\pm 0.03$ &18.8&  7.27&0.75&  0.025&3.31&two-knots&unclear
\enddata
\tablenotetext{a}{Named so that right ascension and declination are implicit.}
\tablenotetext{b}{From the spatially integrated spectrum.}
\tablenotetext{c}{Integrated $g$ magnitude provided by SDSS/DR9.}
\tablenotetext{d}{Masses from SDSS/DR9 photometry using
mass-to-light ratios  by \citet{2001ApJ...550..212B}.}
\tablenotetext{e}{Star Formation Rate from H$\alpha$ flux 
using the prescription in \citet{2012ApJ...750...95E}.}
\tablenotetext{f}{Time to form all stars in the galaxy at the current SFR -- inverse specific SFR.}
\label{the_table}
\end{deluxetable*}

%
\section{Spatial variation of metallicity}\label{analysis}


The direct-method based oxygen abundance corresponding to the 
spatially
integrated spectra of all the observed galaxies is given in 
Table~\ref{the_table}. They represent the luminosity-weighted 
average metallicity. Even though all targets are metal-poor, 
only {\tt J2238+14} is XMP in the usual sense of having 
an average metallicity smaller than a tenth of 
the solar value 
\citep[i.e., $12+\log({\rm O/H})\leq 7.69$;][and Section~\ref{introduction}]{2009ARA&A..47..481A}.

\begin{figure}
\begin{center}
\includegraphics[width=0.35\textwidth]{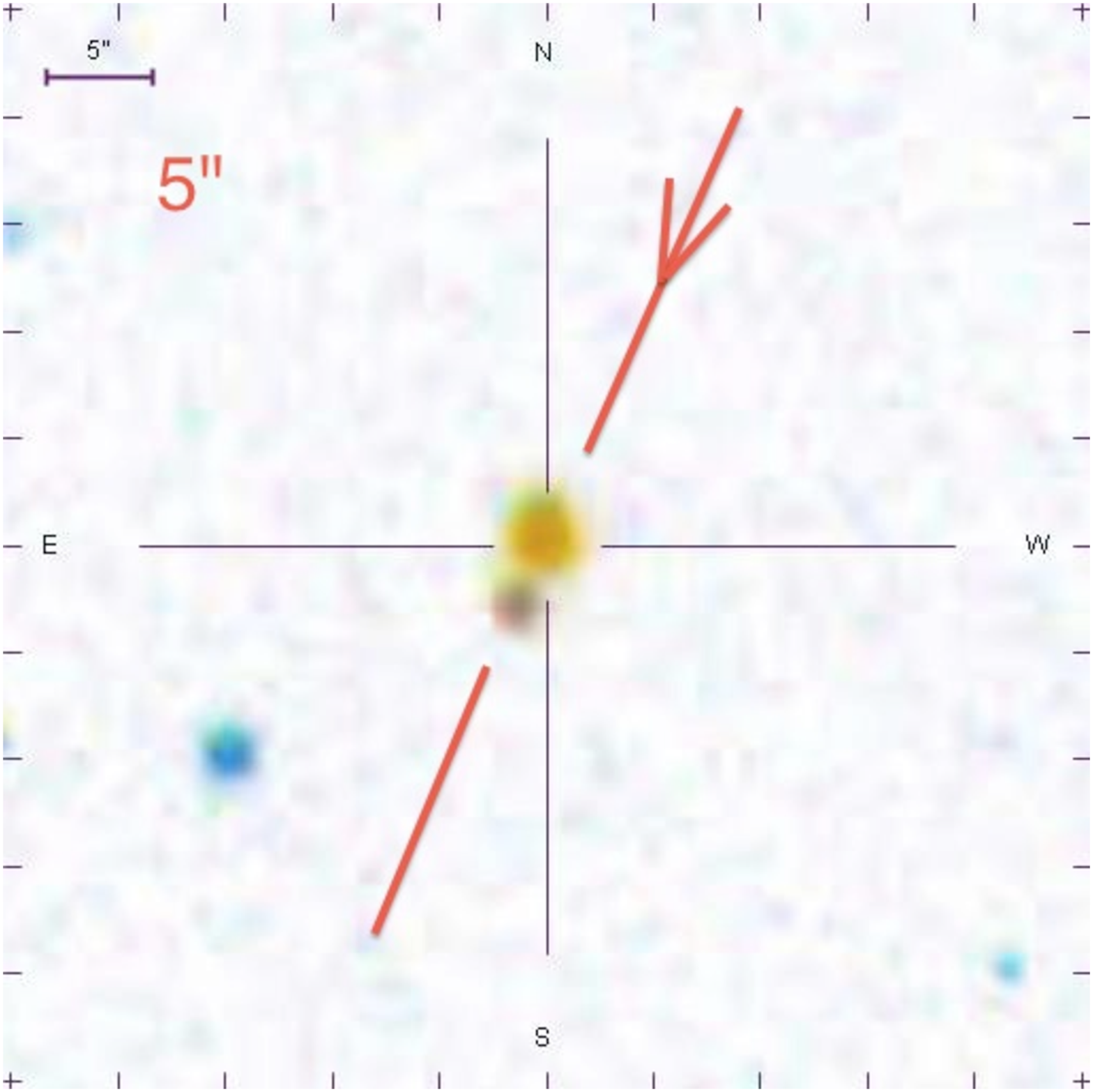}
\end{center}
\includegraphics[width=0.6\textwidth]{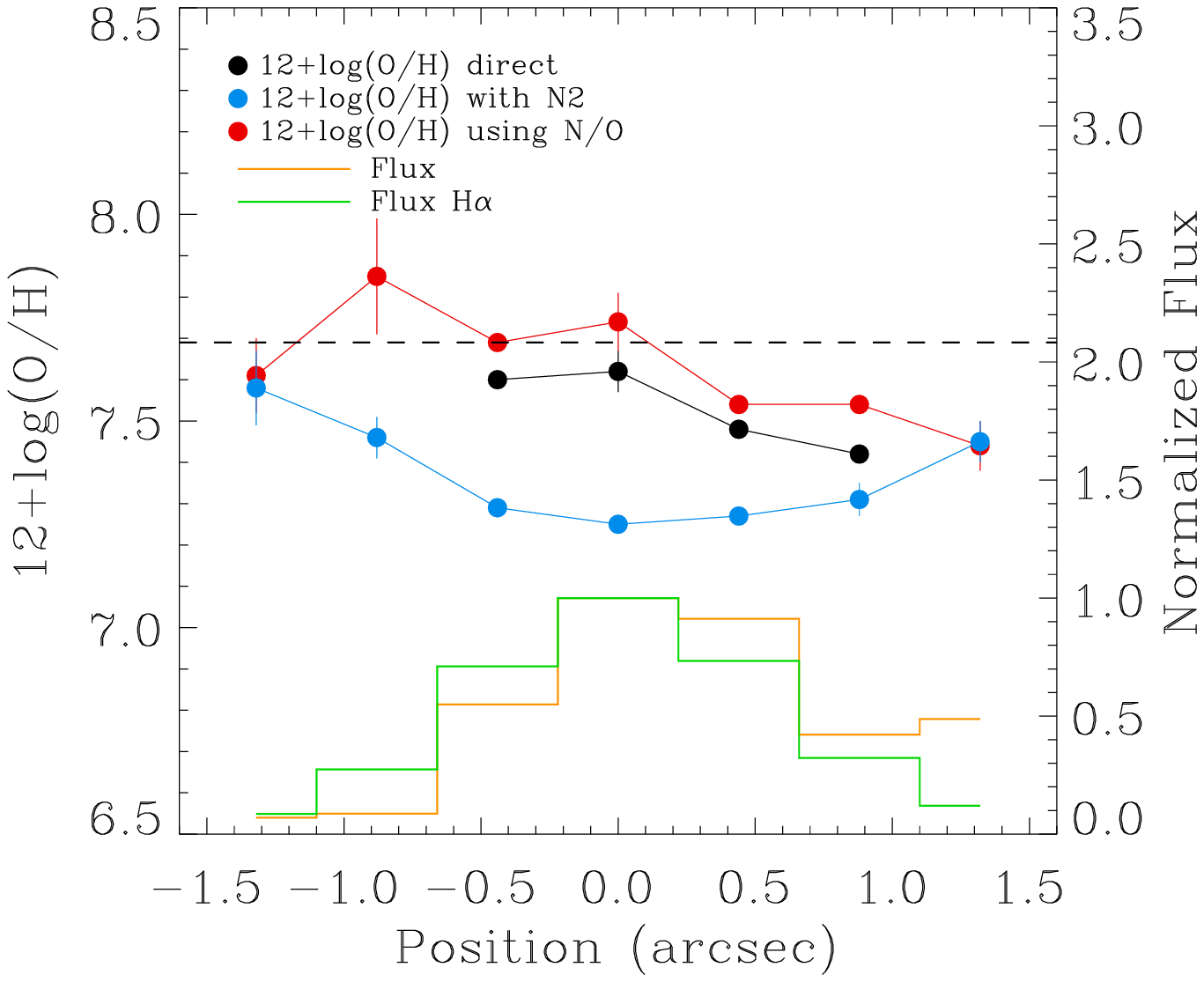}
\caption{
Top: color-code inverted SDSS image of the galaxy  {\tt J2302+00}.
The red line indicates the orientation of the spectrograph 
slit, with the arrow pointing in the sense of growing position
along the slit. The scale corresponds to 5\arcsec on the sky.
Bottom: metallicity and flux variation along the slit of this
target. 
We show the metallicity computed using the direct 
method (the black solid line joining black points), the modified direct method 
described in Sect.~\ref{metallicities} (the red solid line), and
the N2 method (the blue solid line).  
The flux of the integrated  spectrum and the H$\alpha$ flux are given as 
orange and green histograms, respectively. 
Their values have been normalized 
to the largest flux, and the ordinate axis on the 
right-hand-side of the plot refers to them.  The scale of oxygen 
metallicity is given on the left-hand-side of the plot. Positions 
along the slit are  in arcsec refereed to the point of largest H$\alpha$
flux.
In this particular target only the main galaxy knot 
is detected. The extent 
is too small as compared to seeing 
to decide whether there are significant variations  of metallicity. 
The dashed line indicates a tenth of the solar metallicity, a line 
used for reference.
The error bars account only for random
noise in the observed spectra.  
Other sources of error are included in  Fig.~\ref{second_one}. 
}
\label{third_one}
\end{figure}

\begin{figure}
\begin{center}
\includegraphics[width=0.35\textwidth]{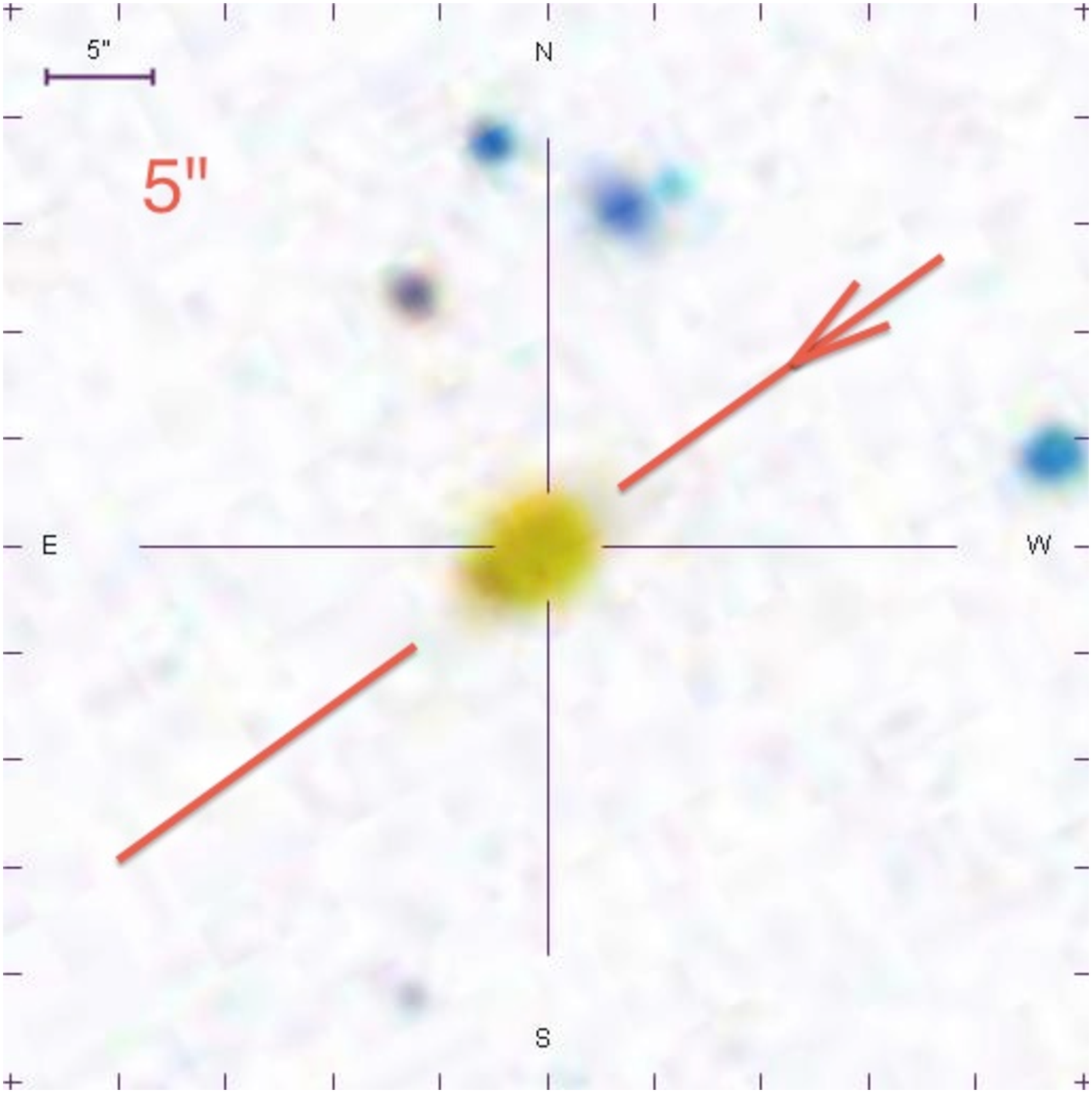}
\end{center}
\includegraphics[width=0.6\textwidth]{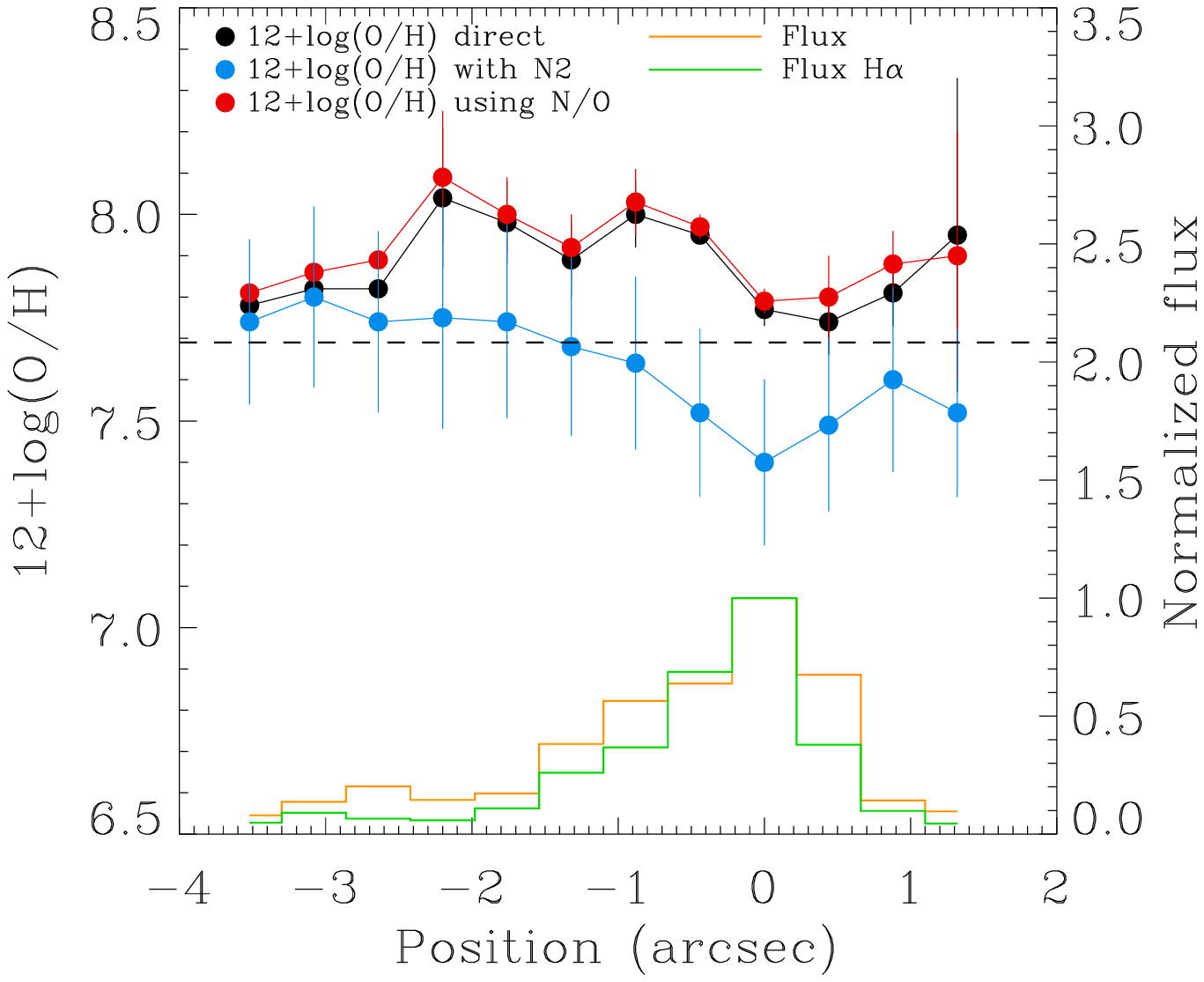}
\caption{
Same as Fig.~\ref{third_one} but corresponding to the target {\tt J1509+37}.
This time  
the spectral signals extend over a region 
larger than the seeing and so we detect
no obvious metallicity variation along the slit. For the meaning
of the various axes, curves and symbols, see Fig.~\ref{third_one}.
The dashed line corresponds to 1/10 of the solar metallicity. 
The continuum noise has been artificially increased by a factor
of three to compute the error bars in this plot.
In addition, the N2 abundance errors have been
enlarged by 0.2\,dex to include  the scatter of 
the N2 metallicity calibration.
}
\label{second_one}
\end{figure}

\begin{figure}
\begin{center}
\includegraphics[width=0.35\textwidth]{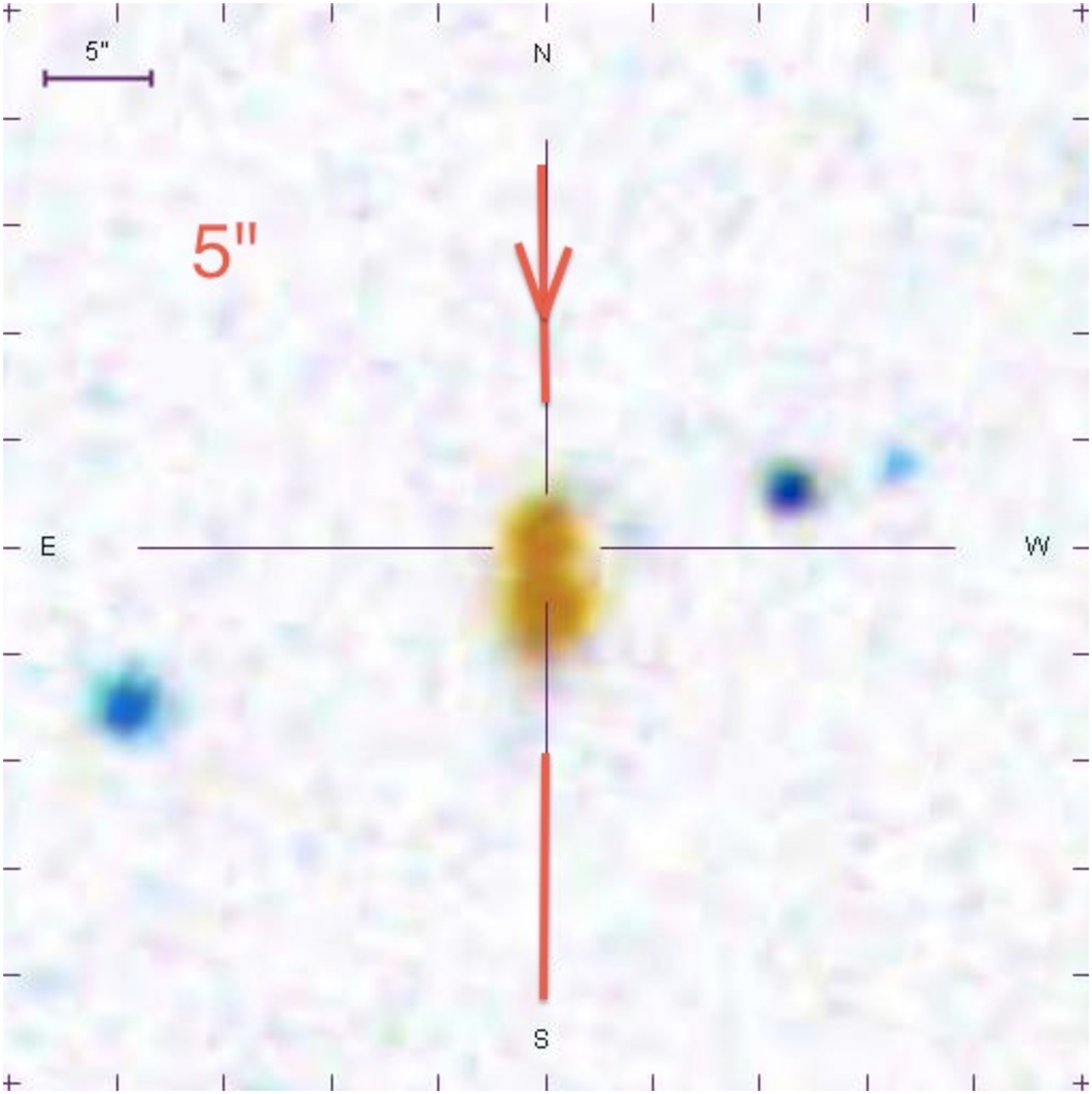}
\end{center}
\includegraphics[width=0.6\textwidth]{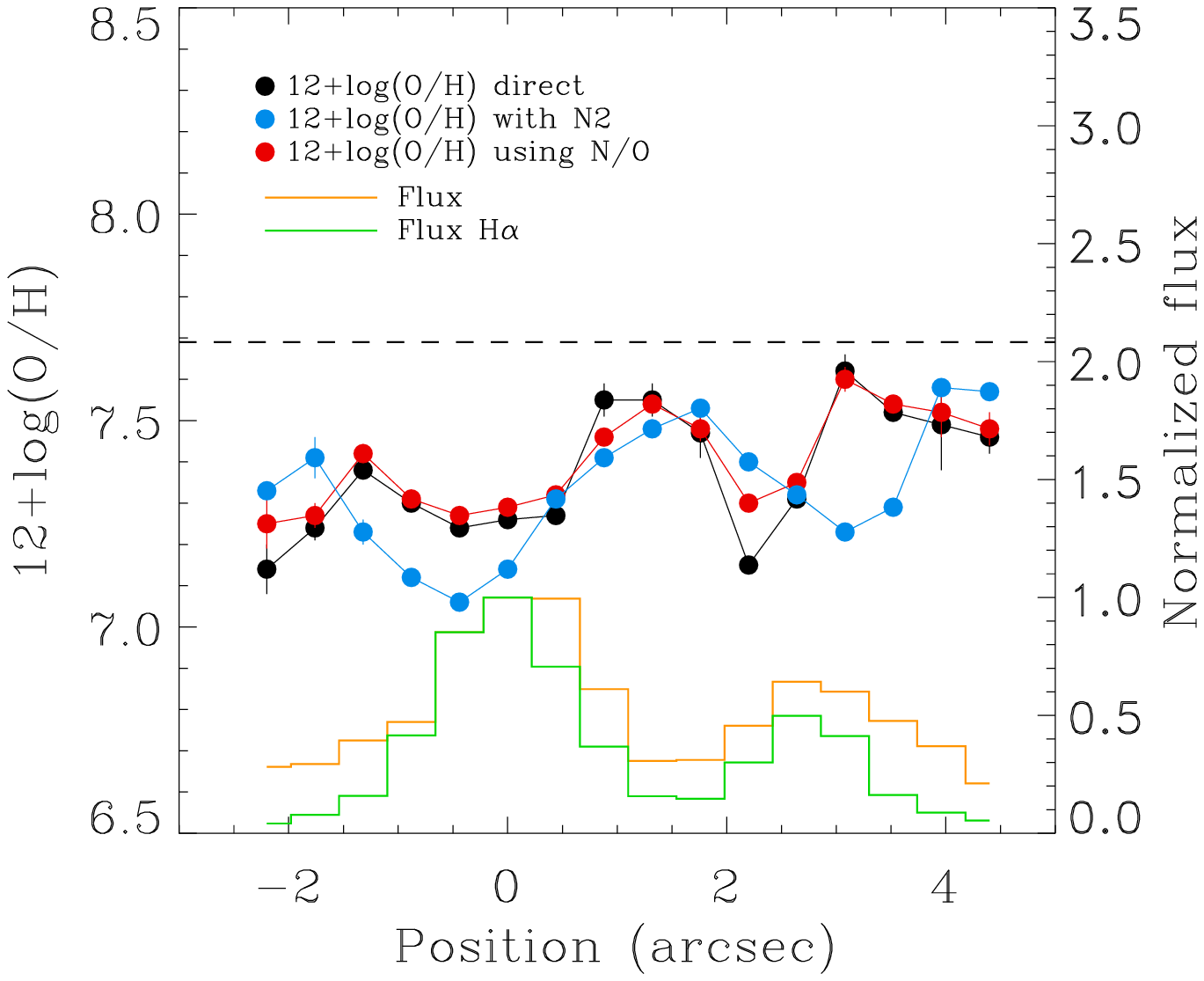}
\caption{
Same as Fig.~\ref{third_one} but corresponding to {\tt J2238+14}.
The target is larger than the seeing, and it shows a clear
metallicity variation along the slit (the black solid line)
with a pattern  similar to the H$\alpha$ flux variation (the green 
histogram). 
For the meaning of the other curves and symbols, see 
Fig.~\ref{third_one}.
The dashed line corresponds to 1/10 of the solar metallicity. 
The error bars account only for random
noise in the observed spectra.  
Other sources of error are included in  Fig.~\ref{second_one}. 
}
\label{first_one}
\end{figure}

\begin{figure}
\begin{center}
\includegraphics[width=0.35\textwidth]{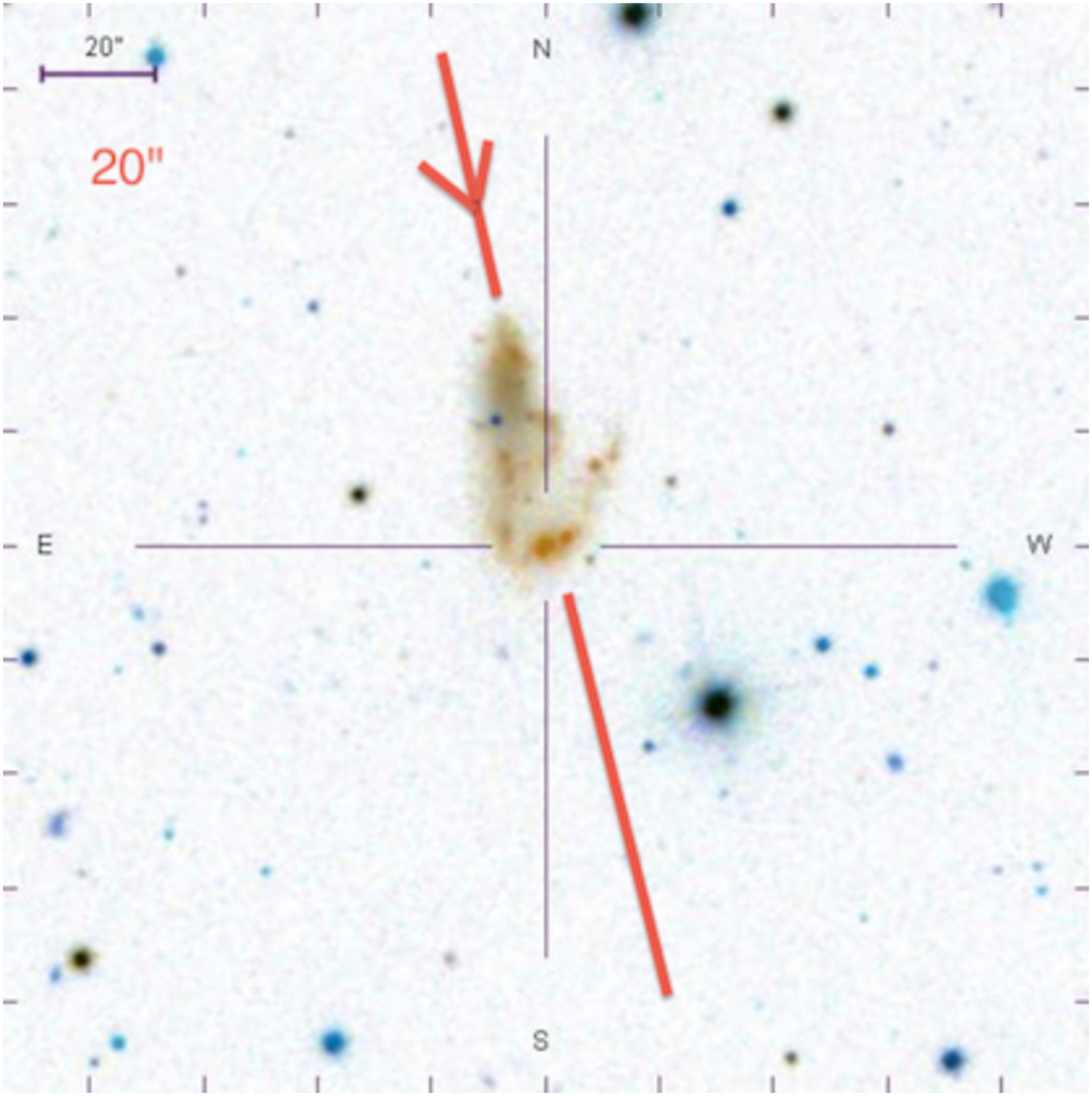}
\end{center}
\includegraphics[width=0.6 \textwidth]{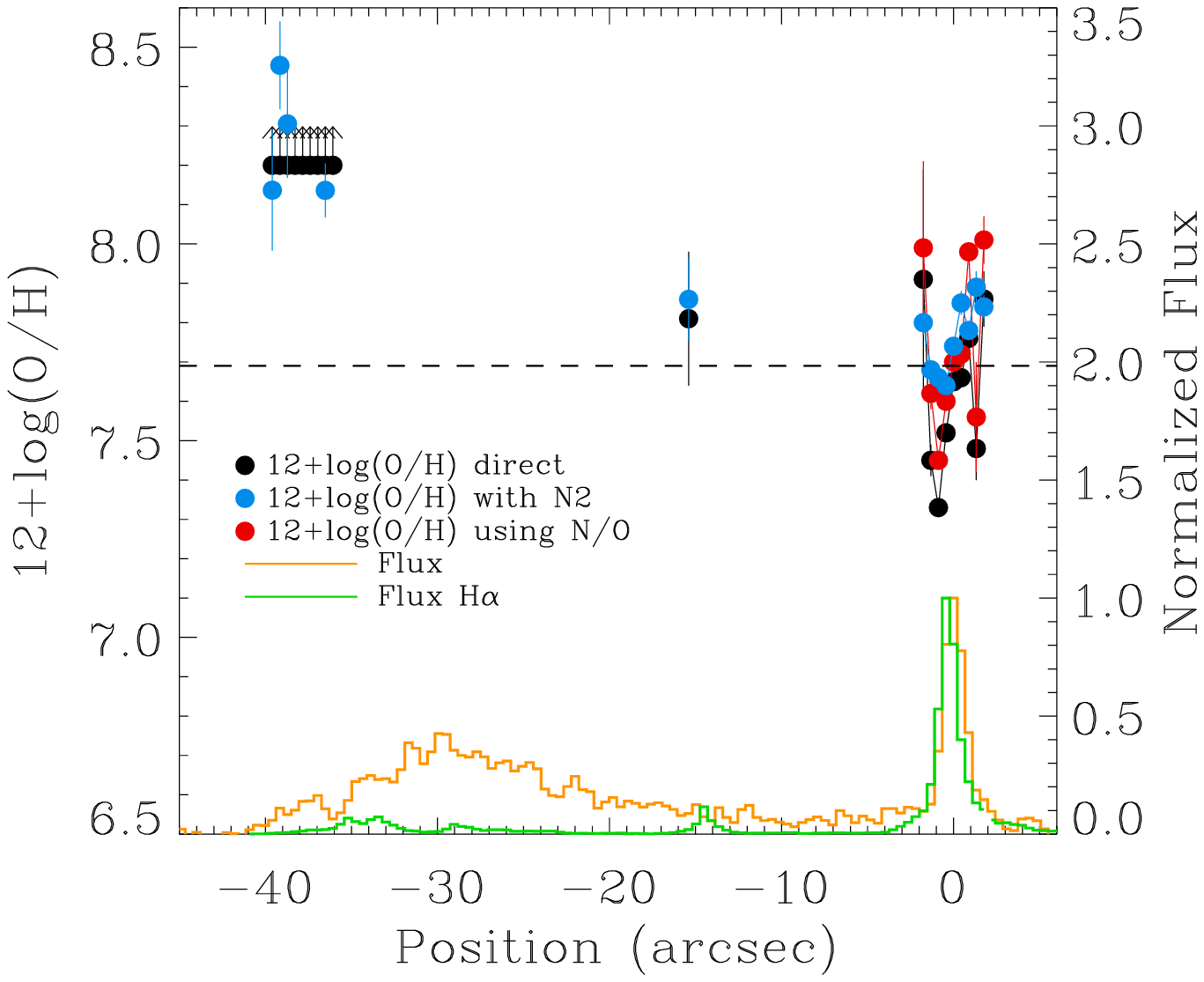}
\caption{
Same as Fig.~\ref{third_one} but corresponding to 
{\tt J1647+21}
The galaxy is larger than the seeing, and it shows
metallicity variations along the slit (the black dots),
with a pattern similar similar to the H$\alpha$ flux 
variation (the green histograms). 
For the meaning of the other curves and symbols, see 
Fig.~\ref{third_one}.
The zero of the position scale in the bottom plot corresponds
to the center of the large cross in the upper image. Positions
along the slit
grow in the sense indicated by the arrow on the top image. 
The error bars account only for random
noise in the observed spectra.  
Other sources of error are included in  Fig.~\ref{second_one}. 
}
\label{extrange_case}
\end{figure}

Figures~\ref{third_one}, \ref{second_one} and  \ref{first_one}
show the three types of observed spatial variations. The Sloan Digital Sky Survey 
\citep[SDSS][]{2002AJ....123..485S,2012ApJS..203...21A}
images on top indicate the orientation of the slit. 
The bottom panels   
plot oxygen abundance versus position along the slit
in arcsec, using as reference position the pixel of largest
H$\alpha$ flux. 
Abundances inferred from the direct method are represented
as black dots joined by black solid lines. These are the measurements we discuss unless
otherwise stated.
Abundances from the modified direct method (the blue lines) 
and N2 (the red lines) are analyzed later on.
The targets {\tt J0942+34, J2302+00} behave similarly, 
the latter represented in Fig.~\ref{third_one}, in the 
sense that the spatial region with enough S/N to carry out the 
metallicity measurement is too small to provide any reliable 
spatial variation. Seeing during observation was of the order of 1\arcsec\ 
(Sect.~\ref{observations}), which is similar to the spatial extent
of the signals on the CCDs (see the continuum and H$\alpha$ fluxes
in Fig.~\ref{third_one},
represented as the orange and green histograms, respectively).
Figure~\ref{second_one}, the black solid line, shows 
a rather constant metallicity, and this time the galaxy is significantly 
larger than the seeing.  The figure displays  
{\tt J1509+37}, but its behavior also stands for {\tt J1003+45}.
These galaxies show no obvious metallicity gradient or drop. 
Finally, Fig.~\ref{first_one} portrays  {\tt J2238+14}
which clearly shows two metallicity decrements associated with the
two bright knots of the galaxy -- compare the black solid
line representing O/H with the H$\alpha$ and continuum fluxes
shown as histograms. The metallicity drop corresponds to
$\Delta[12+\log({\rm O/H})]\simeq -0.5$.  
{\tt J2238+14} is the galaxy of lowest average 
metallicity in the sample
(see the dashed line corresponding  to 
${\rm O/H}=({\rm O/H})_\odot/10$, 
which is common to the three figures).
Figure~\ref{extrange_case} is similar to Fig.~\ref{first_one}
in the sense of
showing a significant spatial variation of metallicity for {\tt J1647+21}. 
The source is larger and more complex
than {\tt J2238+14} (Fig.~\ref{extrange_case}).
The long slit spectrum has not enough
signal for metallicity analysis in between the two main galaxy 
knots (except for a single pixel in between; 
see Fig.~\ref{extrange_case}). However, the signals in the knots clearly indicate a
significant difference of metallicity. 
The brightest one (at position zero)  
has a metallicity of the order of 1/10 solar, whereas the 
second one (at positions between $-35$\arcsec and $-40$\arcsec)
is doubtless metal richer, even though we cannot assess its
actual metallicity. The spectrum in the 
metal-rich knot  does not show [OIII]$\lambda$4363\,\AA ,
needed for electron temperature estimates.
However, this lack 
implies a low electron temperature, and so a high metallicity
-- \citet[][]{1991ApJ...380..140M,2012ApJ...756..163S}.
We estimate a lower limit of  $12+\log({\rm O/H})\geq 8.2$ 
computing the metallicity with a  [OIII]$\lambda$4363\,\AA\ flux 
just below the continuum noise level in our spectra.
This lower limit is represented in Fig.~\ref{extrange_case}.

Table~\ref{the_table} contains a  flag indicating whether the 
metallicity  variations are present in the galaxies, 
are not present, or are unclear. It is unclear in the two galaxies
that are too small. Discarding them, 40\%  of the 
galaxies show metallicity drops (two out of five objects).  

Figures~\ref{third_one}, \ref{second_one}, \ref{first_one}
and \ref{extrange_case}
include metallicities derived from the semi-empirical
N2 method (the blue dots joined by blue solid lines).
Overall, they show the same trends and drops
as the direct method confirming that at least 
for these targets both techniques provide qualitatively consistent
results. 
In some cases there are small differences, e.g., 
the drop of N2 metallicity in {\tt J1509+37}, which is not
obvious in  the direct method based metallicity  (Fig.~\ref{second_one}).
However, these discrepancies are within the 0.2\,dex scatter typical 
of the N2 calibration 
\citep[e.g.,][]{2004MNRAS.348L..59P,2009MNRAS.398..949P}.
This is more clear in Fig.~\ref{second_one} where the error bars
include both the noise in the spectra plus  0.2~dex 
ascribed to the N2 calibration.  
The figures also include oxygen abundances inferred from 
the alternative direct method (Sect.~\ref{metallicities}),
and they also agree with the rest (see the red solid lines
in the figures).
The error bars propagated from the noise
in the spectra  (Sect.~\ref{metallicities}) are  
unrealistically small for reasons that we ascribed
to bias in the flux estimates not accounted
for when propagating the continuum error.
In order to make them
more realistic,  errors were computed also increasing
the observed continuum noise by a factor of three.
These other larger error bars are only represented 
in Fig.~\ref{second_one}.

The galaxy {\tt  J0837+36} has not been mentioned 
so far because its case  slightly differs from the rest. It is more massive 
and with lower specific SFR  (i.e., SFR/${\rm M}_\star$; Table~\ref{the_table}). 
Its metallicity is so large that [O{\sc iii}]$\lambda$4363\,\AA\  
is not detectable in individual spatial pixels and so, 
direct-method metallicity gradients cannot  be computed.
As we explained above for the high metallicity 
knot of  {\tt J1647+21}, the absence of this line proves 
the high metallicity of the HII gas, even though we cannot 
quantify it. From the N2-based estimate,  we conclude that the 
metallicity variations of this galaxy are negligible 
small, as indicated in the last  column  of 
Table~\ref{the_table}.


%
%
%
%
\section{Alternatives to the inferred metallicity 
decrements}\label{rule_out}
This section analyzes alternatives to explaining the 
observed metallicity drops as the outcome 
of  a metal-poor gas accretion event. Specifically, we point 
out possible biases of the direct method that artificially produce
low  metallicities  (Sect.~\ref{metallicities}),
as well as a mechanism that may lead to the metal
impoverishment of regions with long-lasting intense 
starbursts.
Even though these potential problems cannot be fully  
ruled out, accretion of pristine gas remains to be
the simplest way of explaining the observations.

If the temperature in the region is not homogeneous, then
the direct method underestimates the true abundances
\citep{1967ApJ...150..825P,1969BOTT....5....3P}. 
There is a long-lasting  debate in the
literature on whether such temperature fluctuation exists 
\citep[][and references therein]{2013A&A...551A..82S}. 
An artificial reduction of metallicity of 0.5~dex can be 
produced by temperature fluctuations of the order of 
20-30\% rms \citep[e.g.,][]{2004cmpe.conf..115S,2009ApJ...700..654E}. 
If this effect is responsible for the observed     
drop of metallicity, then such fluctuations should be 
localized in the low metallicity starburst, but not in the rest
of the galaxy. We cannot rule out this possibility 
since the interstellar medium (ISM) of our targets is 
poorly known.  We note,
however, that the physical mechanisms proposed to generate
temperature inhomogeneities  favor high metallicity media
rather than our XMP galaxies
\citep[see the review by][]{2003IAUS..209..363T}.
For example, they  require dust particles for 
heating \citep[e.g.,][]{2001A&A...379.1024S},  or they 
need metals for the metallicity inhomogeneities to cause 
the temperature fluctuations \citep[e.g.,][]{1998ApJ...506..323K}.

The direct method does not consider the presence of 
density fluctuations, which in some real cases may be 
large. The resulting electron density variations are not
expected to have significant impact on the abundances
\citep[e.g.,][]{2012EAS....54....3S},
however, density inhomogeneities may have an indirect influence 
though induced temperature inhomogeneities 
\citep[e.g.,][]{2003IAUS..209..363T}. 
If the plasma is heated by collisions with photo-ionized 
electrons, dense clumps present lower temperatures, 
leading to temperature fluctuations. 
Detailed tailored modeling is required to assess
the practical importance of the effect. If it is meant to 
explain the observed metallicity drops, the largest 
density fluctuations must occur where the metallicity
appears to be lowest.

In order to determine the oxygen abundance, 
the standard direct method used in the paper includes
only  O, O$^+$ and O$^{2+}$, but not higher ionization
states. Since the ionizing radiation field is harder in 
young star-forming regions, one may wonder whether our 
metallicity drops are actually caused by overlooking O$^{3+}$
in large starbursts. However, this potential bias does not 
explain the magnitude of the observed drops.
Even when very hot stars are present in HII regions, the 
correction for  unseen states of oxygen is negligible small with 
respect to other sources  of  errors \citep[e.g.,][]{2012EAS....54....3S}.
The ionization correction factors for O$^{3+}$
are never of the order of 0.5~dex as required to 
reproduce our observations \citep[e.g.,][]{1994MNRAS.271..257K}.

Dwarf galaxies have shallow gravitational potentials that 
cannot retain all the metals ejected by  SNa 
explosions \citep[e.g.,][]{1999ApJ...513..142M}. 
Therefore, their metal enrichment  depends critically on 
two competing processes,  both controlled 
by the SNa rate, that is to say, controlled
by the SFR -- the metal production and the metal loss.
Both increase with the SFR.  Due to this interplay, 
dwarfs may enrich more 
efficiently at mild SFRs, where the two 
opposite effects reach a compromise
\citep[][]{2011ApJ...730...14H,2013MNRAS.428.2949K}.
This tradeoff between SFR and metallicity may induce
the metallicity pattern that we observe.
If a major starburst has been losing most of the 
metals for long, the gas around it would have a metallicity 
lower than the rest of the galaxy, where the star-formation 
has proceeded at a lower more-efficient rate.  
Even though we cannot fully discard this possibility, 
we envisage two difficulties for this explanation to work
with our targets.
First, the high star-formation mode quickly exhausts 
the original gas supply, which has to be replenished
with metal poor gas that does not exist in the galaxy.
Second, and equally important, the gas in the star-bursting
region should remain unmixed or the full galaxy would 
acquire a uniform metallicity.  This is not easy to 
attain since mixing mechanisms are expected  
to efficiently operate in short time-scales 
\citep[Myrs; e.g.,][]{1996AJ....111.1641T,2002ApJ...581.1047D}.

%
%
%
\section{Ubiquity of the phenomenon}\label{ubiquity}

A number of observational properties characterizing large samples 
of  star-forming galaxies can be naturally explained if the 
metallicity drop associated with intense starbursts is a common 
phenomenon.  The inflow of pristine gas provides 
a simple physical unifying mechanism that explains all of them,
even though  often it is not  the only explanation of each individual
observation. This section critically reviews some of these results
in terms of metal-poor gas inflow triggering star-formation. 
The discussions are fairly qualitative, emphasizing the 
diversity of observations hinting at grand-scale 
metal-poor gas inflows in the local galaxies.

\subsection{The mass-metallicity-starformation-rate 
relationship\footnote{Often referred to as {\em fundamental} 
mass-metallicity relationship \citep[][]{2010MNRAS.408.2115M}.
}}\label{fmm_rela}

Local galaxies are known to follow a mass-metallicity relationship,
where the larger the mass the higher the metallicity 
\citep[e.g.,][]{1989ApJ...347..875S,2004ApJ...613..898T,2005MNRAS.362...41G}.
The relationship
presents a significant scatter that has been recently 
found to be associated with the present SFR in the galaxy 
\citep{2010MNRAS.408.2115M,
2010A&A...521L..53L,
2012MNRAS.422..215Y,
2013A&A...549A..25P,
2013ApJ...765..140A,
2013arXiv1310.4950Z}.
Specifically,  for galaxies with the same stellar mass, the 
metallicity decreases as the current SFR increases. 
The mass-metallicity relationship is commonly interpreted as
due to variations of the star-formation efficiency with galaxy 
mass, and/or to galaxy mass-dependent metal-rich outflows
\citep[e.g.,][]{2008IAUS..255..100L,2008ApJ...672L.107E}. 
The former implies that low mass galaxies produce less stars
for their gas, and so become more metal poor, whereas the latter 
relies on the metal-rich SNa ejecta to be preferentially lost to the 
intergalactic medium by low mass galaxies,  due to their 
shallower gravitational well.  
Neither of these two mechanisms, however, 
predict the observed  dependence  on SFR
of the metallicity  -- they render 
a metallicity set only by the galaxy
mass\footnote{
At least in simple chemical evolution models.
Even if the outflow rate scales with the SFR,  
the metallicity of the gas is set by the gas fraction,
but this gas fraction depends only on the stellar mass
\citep[e.g.,][]{1990MNRAS.246..678E}. 
Thus, given the stellar mass, the metallicity is fixed, leaving 
no room for a SFR--metallicity correlation. 
}.
Conversely, the observed anti-correlation between metallicity
and SFR  can be qualitatively understood if  the star-formation is 
preferentially triggered and sustained
by the inflow of metal-poor gas, which has no time to be
well mixed 
with the high metallicity gas already present in the ISM of the galaxies.
The agreement is more than qualitative according to  
\citet{2012ApJ...750..142B}. Using a toy model for 
the gas inflow, these authors conclude that most of the  
star-forming galaxies  with stellar masses 
$M_\star  \leq 2 \times  10^{10}$\,M$_\odot$, 
and many with $M_\star  \geq 2 \times  10^{10}$\,M$_\odot$ 
appear to be fed by low-metallicity gas infall. 
The importance of metal-poor gas infall to account for the 
observed mass-metallicity-SFR relationship is also emphasized by
\citet{2012MNRAS.421...98D}
and \citet{2013MNRAS.430.2891D}
in  their simple analytic 
chemical evolution models that include mass infall and outflows.
In particular,
\citet{2012MNRAS.421...98D} explain the metallicity-SFR 
relationship as transient departures from the secular evolution
of the galaxies,  
triggered by sudden infalls of metal-poor gas.  

%

Unlike the metallicity, the ratio between the observed
N and O does not seem to depend on SFR 
\citep[see][]{2013A&A...549A..25P,2013ApJ...765..140A}.
This lack of SFR-dependence is consistent with the relation 
between metallicity and SFR 
being maintained by episodic metal-poor inflows.
The advent of fresh gas triggers
star formation and drops the metallicity, but it does not
change the pre-existing relative abundance between metals. 

The mass-metallicity- SFR relationship
is followed by large numbers of star-forming galaxies 
so it represents a behavior common to the typical   
galaxies of the local Universe. It 
is not restricted to a few rare vestigial objects. 
Therefore, if the above conjecture turns out to be
correct, and pristine gas infall is responsible for
the SFR dependence, then this infall is a
characteristic of the full population of local
star-forming galaxies. This conclusion 
is the central point of the section.

%
%
\subsection{High metallicity of quiescent BCDs}\label{hi_bcd}

BCD galaxies are high surface brightness 
targets and thus,  relatively easy to detect.  Most XMPs are also 
BCDs \citep[e.g.,][]{2000A&ARv..10....1K,2011ApJ...743...77M}.
The luminosity of these galaxies is dominated by one or 
several young starbursts, however, most if not all BCDs contain host 
galaxies with old stars \citep[e.g.,][]{1996A&AS..120..207P,
2003ApJ...593..312C, 2006ApJ...651..861C,
2007A&A...467..541A}.
The dominant starburst is  so intense that it cannot be sustained 
for long, therefore, the BCDs have to be in a transient phase.
(Using the arguments and symbols in Sect.~\ref{observations}, 
their $t_\star << t_0$.) 
Consequently,  there must be many local galaxies in the 
pre~or~post BCD phase, i.e., many quiescent BCDs 
(or, for short, QBCDs).

The BCD hosts show up  in the galaxy outskirts, therefore,
deep photometry allowed 
\citeauthor{2007A&A...467..541A}~(\citeyear{2007A&A...467..541A}, 
\citeyear{2009A&A...501...75A})
to characterize their photometric properties.
Using the typical host colors and magnitudes as proxies for QBCD
properties, 
\citet{2008ApJ...685..194S}  searched the 
SDSS/DR6 archive for QBCD candidates.
They turned  out to be rather common -- 
one out of three local dwarf  galaxies is of this kind, 
and there are some thirty of them per BCD galaxy. 
Their main properties, including their
luminosity functions, are consistent with the BCDs being 
QBCDs observed during a starburst phase in a duty cycle 
where the QBCD phase lasts 30 times longer than the BCD phase. 
This interpretation presents a difficulty, though: 
the gas-phase metallicity of the QBCDs is systematically 
higher than the metallicity of the BCDs.
This cannot happen in a closed-box
evolution, where the precursor galaxy always has lower
metallicity than the follower, so that QBCDs could not be 
precursors of  BCDs. 
The problem naturally disappears if almost {\em every} 
BCD starburst  is  preceded by the advent of fresh 
metal-poor gas that triggers the star formation episode. 
 Moreover, such 
gas-infall hypothesis beautifully explains why the 
{\em stellar} metallicities of BCDs and QBCDs agree,
even though their {\em gas-phase} metallicities do not 
\citep{2009ApJ...698.1497S}. The stars of BCDs and 
QBCDs are statistically the same because only 
a small fraction of galaxy stellar mass is produced 
in each starburst\footnote{
Although in some extreme cases the present burst may be 
producing a significant fraction of the stellar mass, e.g.,
our  {\tt J1509+37}, this is not the general behavior. 
}.
Their gas differs because BCDs have just 
rejuvenated their ISM. 

This behavior affects not just a few objects, but 30\,\% of all local dwarfs.
Therefore the gas-infall must be a common phenomenon
if it is responsible for the  metallicity discrepancy between
BCDs and QBCDs.  

\subsection{The morphology metallicity relationship}\label{morpho_rela}

XMP galaxies tend have  cometary or other 
non-symmetric morphologies
\citep{2008A&A...491..113P, 2011ApJ...743...77M,2013A&A...558A..18F}.
Even if surprising, such association 
seems to be the extreme case of a common relationship
between morphology and metallicity followed by the bulk 
of the star-forming galaxies in the local Universe.
\citet{2009ApJ...691.1005R}  parameterize lopsidedness in a 
sample of $\simeq 2.5\times 10^4$ nearby galaxies, and find that 
at fixed mass, the more metal-poor galaxies are more lopsided. 
Whatever process causes lopsidedness, it 
it is also associated with low metallicity gas in
the galaxies. In the case of the XMP, 
the lopsidedness is produced by off-center large HII regions, 
fed by pristine gas accretion either directly or 
indirectly  -- directly if the gas arrives to the disk 
ready to form stars 
\citep[e.g.,][]{
2009Natur.457..451D,
2009ApJ...703..785D},
or indirectly if the gas is accumulated until disk instabilities 
trigger star-formation in regions that must be 
necessarily large compared to the disk extension 
\citep[e.g.,][]{
1999ApJ...514...77N,
2008ApJ...688...67E,
2012ApJ...747..105E}. 
Low metallicity and lopsidedness come together naturally
in XMPs. If the physical mechanism that gives rise to the 
cometary shape of XMPs is also responsible for the 
correlation between morphology and metallicity found by 
\citet[][as an Occam's razor type of argument suggests]{2009ApJ...691.1005R}, 
then triggering star formation 
by gas inflow must be quite common.

\subsection{Nitrogen and Oxygen in green-pea galaxies}\label{green_peas}

{\em Green peas} (GPs) are star-forming galaxies which receive this name 
because of their compactness and green color in SDSS composite  images
\citep[][]{2009MNRAS.399.1191C}. 
The color is produced by  an unusually large [OIII]$\lambda$5007\,\AA\ emission 
line redshifted so as to contribute to the $g$-band color. 
They have some of the highest specific SFRs seen in the  
local Universe, able to double their stellar masses in a fraction of Gyr. 
GPs seem to be high-mass versions of the most extreme starbursting 
BCDs \citep[e.g.,][]{2011ApJ...728..161I,2012ApJ...749..185A}, and are  
low metallicity outliers of the mass metallicity relationship
\citep{2010ApJ...715L.128A,2012ApJ...749..185A}.
Detailed analysis of their emission lines reveals complex kinematical 
structures with several components coexisting in only 
a few kpc,  which are best interpreted as massive star-forming clumps 
in a dynamically young host galaxy \citep[][]{2012ApJ...754L..22A}.
Even though GPs have low O metallicity, they present an
overabundance of N/O$(\geq -1)$, which is typical
of aging stellar populations. This puzzling observation is 
naturally explained if GPs have recently 
received a major flood of low metallicity
gas \citep{2010ApJ...715L.128A,2012ApJ...749..185A} --
the mixing with metal poor gas reduces the metallicity (i.e., O/H), 
but the ratio between metal species (N/O) remains as in the original 
high metallicity ISM.

Again, GPs are not special but just extreme cases in the 
continuous  sequence of local star-forming galaxies 
\citep[e.g.,][]{2011ApJ...728..161I,2013RMxAC..42..111S}.

%
\subsection{Other hints of  gas accretion}

The literature contains other results that are also
suggestive of star-formation triggered by gas accretion
at a grand-scale.
Some of them are mentioned below.


The neutral gas distribution of the BCD galaxies often shows large 
distortions, with plumes and tails, and other evidences 
of gas inflow or outflow  
\citep[e.g.,][]{1988MNRAS.231P..63B,
1998AJ....116.2363W,
2012A&A...537A..72L,
2012MNRAS.419.1051L,
2013AJ....146...42A}.
Such complex HI morphology appears even in the case of  isolated galaxies
without obvious companions
\citep{2010MNRAS.403..295E}.
The distorted gas around BCDs has all signs of having extremely low 
metallicity, uncontaminated by the ongoing star-formation 
process \citep[e.g.][]{2013A&A...553A..16L,2013A&A...558A..18F},
which suggests that the gas is arriving rather than being
expelled from the galaxy.


Even large nearby spirals show local metallicity inhomogeneities 
that deviate from the main gradient, e.g., M101 
\citep{2013ApJ...766...17L}. The existence of inhomogeneities 
is in tension with theoretical  expectations, which predict a 
virtually uniform distribution as a result of the short mixing 
timescales of the interstellar medium,  on the order of only 
100~Myr
\citep[e.g.,][]{1995A&A...294..432R,
1996AJ....111.1641T,
2002ApJ...581.1047D}.
Localized infall of metal-poor gas may be  a viable alternative 
that explains this particular observation.

As we pointed out in connection with BCDs and GPs 
(Sects.~\ref{hi_bcd} and \ref{green_peas}), the gas inflow 
produces large excursions of a galaxy in the N/O vs O/H plane.
Numerical models by \citet{2005A&A...434..531K} 
allow to explain the observed distribution in irregular
and spiral galaxies, but, in order to reach the required 
large excursions, 
the mass of the infall gas must be much larger than the mass 
of the gas present in the galaxy, with the infall rate  exceeding the SFR.

This trend for the low metallicity galaxies to show 
anomalous metallicity gradients (Sect.~\ref{analysis}) 
is also observed  at high-redshift 
\citep[e.g., the z=1.2  
MASSIV  galaxies; ][]{2012A&A...539A..93Q}.
and in several low redshift targets 
\citep[e.g.,][]{2011ApJ...739...23L,2010ApJ...715..656W}.
The metal-poor galaxies tend to show a positive gradient, 
whereas metal-rich ones tend to show the negative one
expected from secular evolution.
Positive gradients naturally arise even from underlying negative 
gradients when metal-poor gas reaches the central regions of the disks.

%
\section{Discussion and conclusions}\label{discussion}

We measure the oxygen metallicity along the major axis 
of seven star-forming dwarf galaxies using 
different methods,
including the direct method (Sect.~\ref{metallicities}; Table~\ref{the_table}).  
Two of them,   {\tt J1647+21} and {\tt J2238+14}, 
show drops of metallicity ($\simeq$0.5\,dex) 
associated with enhanced star-formation 
activity in central regions.   
Disk galaxies usually present a negative gradient, with the metallicity 
decreasing inside out. Therefore, a deficit of metallicity in the inner 
galaxy is strange, and attributed to the recent arrival 
of external metal-poor gas that has not yet mixed up with 
the pre-existing ISM (Sect.~\ref{introduction}; 
other alternatives are also examined in Sect.~\ref{rule_out}).
For this to happen,
the incoming gas has to arrive in localized 
clumps rather than as an isotropic galaxy-wide accretion 
event.
This is the explanation we suggest for the metallicity and H$\alpha$
variations observed in  {\tt J2238+14} (Fig.~\ref{first_one}), and
{\tt J1647+21} too  (Fig.~\ref{extrange_case}).
The image of the latter, however, may also suggest 
a merger event, with the main starburst at the collision  
point (see  Fig.~\ref{extrange_case}, with the two colliding 
disks seen edge-on forming a V-shape in a contrived but not
impossible geometry). One of the galaxies would have
to be metal-poor gas-rich, with its gas feeding 
the low metallicity starburst. Actually, such a gas-rich minor
merger can also be regarded as a cold-flow accretion event
where the accreted gas stream is forming stars along the way
\citep[see][]{2009Natur.457..451D}.
The different morphology of {\tt J2238+14} and
{\tt J1647+21} may be due to differences in  
spatial resolution, so that we have a coarser
view of  the former. But we cannot discard that they reflect 
qualitative differences in the physical process 
responsible for the metallicity drops.

Our interpretation of the metallicity drops of  {\tt J1647+21} 
and {\tt J2238+14}  agrees with that
given by \citet{2013ApJ...767...74S} to explain the behavior 
observed  in a number of local tadpole galaxies.
Such agreement
has several implications. 
It proves that
galaxies other than the sample of tadpole galaxies 
\citep{2010PNAOJ..13....9M,2012ApJ...750...95E}
present the same unusual  spatial metallicity pattern. 
The metallicity inhomogeneities of our targets
were inferred using the direct method, which discards 
the systematic errors usually attributed to 
strong-line methods 
\citep[e.g.,][]{2005A&A...437..849S,2009MNRAS.398..949P}.
This source of error is discarded for  {\tt J1647+21} 
and {\tt J2238+14}, thus supporting the type
of metallicity pattern disclosed in tadpoles  
\citep{2013ApJ...767...74S}.
Finally, the targets showing the drops tend to have 
a minimum metallicity smaller than a tenth of the solar 
value. Something similar
happens with the tadpoles 
analyzed by  \citet{2013ApJ...767...74S}, which may suggest
a tenth of the solar metallicity to be an observational
threshold for the metallicity drops to clearly show up.
The origin of the threshold is unclear, but it may  reveal an  
observational bias reflexing the degree of mixing of the 
galaxy gas.
Assume that all 
typical disks have similar fairly high gas metallicity, and they 
receive a parcel of metal poor gas. Those galaxies that mix up this 
metal poor gas with the pre-existing ISM before star-bursting 
will appear as metal rich targets of homogeneous metallicity. 
On the contrary, those that produce stars before mixing
will look like metal poor galaxies in integrated  
light, presenting large metallicity inhomogeneities.
The actual threshold is probably not universal since 
high redshift galaxies with drops have metallicities
above the one tenth line \cite[][]{2010Natur.467..811C}.

If  external metal poor gas accretion feeds and triggers 
star formation, one would expect some kind of kinematical 
differences between the star-forming clumps and the 
underlying  galaxy disk. These  kinematical disturbances are predicted in
numerical  simulations of minor mergers and cold-flow accretion  
\citep[e.g.,][]{2004ApJ...611...20I,2009ApJ...703..785D,2010MNRAS.404.2151C},
and should be  sought in real galaxies.
Some of then may have been observed already as,
e.g., the counter-rotating head found in one of the tadpole
galaxies analyzed by \citet{2013ApJ...767...74S}.

Large metallicity inhomogeneities evidence  a star formation  driven, 
or at least stimulated,  by pristine gas  accretion. Even though 
the number of local galaxies showing 
inner metallicity inhomogeneities is still limited,
there are a number of indirect hints  suggesting
that metal-poor gas accretion may be more than just a 
vestige of the early Universe. The argument relies on 
the existence of general rules or laws followed by large numbers 
of galaxies, that are naturally explained as star-formation
triggered by recent pristine gas infall. It
is not the only explanation, but the inflow of pristine gas 
provides  a simple unifying physical  mechanism that explain all of them.
These evidences are outlined in Sect.~\ref{ubiquity}: among others, 
the star-formation dependence of the metallicity (Sect.~\ref{fmm_rela}), 
the star-formation dependence of the morphology (Sect.~\ref{morpho_rela}), 
the high metallicity of quiescent BCDs (Sect.~\ref{hi_bcd}),  and
the high N to O ratio in green pea galaxies (Sect.~\ref{green_peas}).

%

\begin{acknowledgements}
Thanks are due to R.\,Amor\'\i n and  E.\,P\'erez-Montero
for enlightenning discussions on the constraints on 
galaxy evolution provided by the observable N/O,
%
to V.~Luridiana and M.~Koleva for clarifying some of 
the biases of the direct method, and
to J.\,A.\,L.\,Aguerri for technical 
support during the observing campaigns.
Thanks are also due 
an anonymous referee for helping us sharpening 
some of the arguments in the paper.
This work has been partly funded by the Spanish Ministry for Science, 
project AYA~2010-21887-C04-04.          
JMA acknowledges support from the European Research Council Starting Grant (SEDmorph; P.I. V. Wild). 
The article is based on observations made with the telescope WHT  
operated at the Spanish {\em Observatorio del Roque de los Muchachos} 
of the {Instituto de Astrof\'\i sica de Canarias}.
%
\end{acknowledgements}
%
%
\newcommand\nar{New Astron. Rev.}
\bibliographystyle{aa}

\end{document}